\documentclass{article}[12pt]
\usepackage{amsfonts}
\usepackage{eucal}
\usepackage{amssymb}
\usepackage{amsmath}
\usepackage[T1]{fontenc}
\newcommand{\rmd}{{\rm d}}

\begin{document}

\title{The Production of Time}

\author{Adam Helfer}

\maketitle

The nature of time is arguably the most fundamental question in
physics today.  It includes the daily mystery of reconciling the
perceived unfolding of events with the apparently almost entirely
reversible laws of the physics which successfully describes so much of
them, the puzzle of the very existence (to good approximation) of a
``cosmic time,''
the question of what replaces conventional notions of time at
Planck scales near the origin of the Universe (and so the question of
in what sense the Universe {\em had} an origin, and whether there is
any meaning to what happened ``before the origin''), as well as the
ultimate end of the Universe.

We have no clear indication at present that we are close to really
understanding any of these points.  What I shall do here is to explore
some of them from the perspectives offered by attempting to
reconcile two successful and (from our present view) fundamental
theories, general relativity and quantum field theory.  I shall argue
that there is good reason to think that the measurement process and
issues related to it are bound closely with understanding the nature
of time, and, indeed that measurement 
creates ``time'' in a sense which is as important
as the usual relativistic understanding of time.

\section{General Relativity}

I think most physicists would agree that general relativity coupled to
appropriate classical sources gives a very beautiful and satisfying
picture of the development of systems over time, in the sense that the
space--time resulting from such a description can be relied on within
well-understood limits to describe physical reality, and contains
within it a prescription for understanding and measuring time:
insofar as general relativity is applicable, time is measured by
the geometry of space--time itself, the ``proper time'' along causal
curves.

On the other hand, the matter terms chosen to be sources for general
relativity can hold within themselves physics which gives one pause
about the completeness of such a classical general-relativistic
picture.  The most vivid of these are contained in ``time bomb''
scenarios:  situations where the matter, at some point, because of an
infinitesimal internal change, winds up altering its behavior and, in
turn, setting off enormous changes in the structure of space--time
itself.  In this way, depending on whether the ``bomb'' goes off or
not, one can arrange for galaxies to collide or not, for black holes
to form or not, and for the entire causal structure of space--time
(including not just quantitative measures of time but the relative
temporal orderings of events) to
the future of the ``explosion,'' to be subject to the behavior of the
``bomb.''

One might take the position that such situations are self-consistently
described by classical physics; they simply depend sensitively on
initial conditions (the precise time the bomb will go off, the
precise nature of the explosion).  Yet it is not hard to arrange that
these classical uncertainties are dominated by quantum ones.  Then,
whatever the classical model's self-consistency, it neglects essential
physical effects and is not appropriate for predictive
work.  In a case like this, one can
presumably {\em describe} and {\em measure} the evolution of the
space--time and thus of measures of time within it
by classical constructs, one does not
have an adequate way of {\em predicting} it from initial data, in more
than a statistical sense.

This is closely related to a discrepancy between how general
relativity describes the world and how we experience it.  We have a
clear sense of existing in the present, with a future which has yet to
be created.  And while general relativity does an excellent job of
describing the course of events and their causal relations once they
have transpired, it does not, except for restricted forms of classical
matter, give an absolute way of predicting the future.  

It is no accident that these potential failures of predictability seem
to depend essentially on properties of the matter terms, rather than
on more purely gravitational factors.  This reflects the common view
that some form of ``cosmic censorship'' should hold and hence that in
some sense the maximal Cauchy development of the space--time should be
inextensible; were cosmic censorship to fail, one would have more
essentially gravitational sources of unpredictability.

\section{Quantum Measurement}

I want to turn now to the quantum realm.  I will adopt the Heisenberg
picture, so that state vectors do not change (except when they are
reduced by measurements --- an issue I will take up soon), 
and the quantum operators evolve.  There has, of course, been a great
deal of debate about different possible interpretations of quantum
theory.  I would like to focus this by beginning with a physical
question.  (I will leave it to my readers to judge how well different
interpretations address this question.)

I would suggest that the main unanswered question in quantum theory is, When
does a measurement occur?  This question is not directly answered by any
conventional interpretation (although it has been taken up within the framework
of ``objective reduction'' theories, cf.~Leggett 2002).  More precisely, under
what circumstances can we say that a measurement will occur, or will be likely
to occur, and what observable will be measured, or be likely to be measured?

While in a practical sense we answer this in laboratories all the
time, that practical answer seems to come down to a matter of human
consciousness --- of reading a value.  While some have, on this basis,
argued that consciousness is an essential part of the reduction
procedure, I find this very hard to accept.  (Is it really plausible
that the Universe was in a gross macroscopic superposition of states
until consciousness developed?  And, what started
consciousness?)  Setting explanations via
consciousness aside, then, we have at present little to say in answer
to this question.\footnote{I should mention one point at which 
this question is believed to be
especially important, and on which there is an approach to it in contemporary
work.  This is the question of when quantum fluctuations in the
early Universe ``freeze out'' and become classical; cf. Peacock (1999).  
While there is in
the literature a definite prescription for this in terms of ``horizon
crossing'' in particular cosmological models, and while this
prescription may well turn out to have merit, it should be
emphasized that it really amounts to a new hypothesis
and is not a consequence of any accepted physical theory.}

One reason that this question is hard is that there is a great deal of
freedom in ``placing the classical/quantum cut.''  To a large extent,
it does not make any practical difference whether we regard an
experiment as done when its results are printed out or when we look at
the print-out.  Within the framework of quantum theory, we may say
that, as long as the observable to be measured commutes with the
Hamiltonian, it makes no difference at what time it is
observed.\footnote{More precisely, the requirement is that the
observable should commute with the Hamiltonian on the space in which
the state vector is known to lie.}  Yet while this invariance is
substantial, it cannot be pushed to the point of asserting that
measurements do not occur.

When a measurement does occur, I shall say that, insofar as
conventional quantum theory is adequate to describe what happens, the
state vector reduces.  I am aware that many working physicists prefer
to avoid this term.  On the other hand, the actual computations that
are done (projecting to the relevant eigenspace of the operator
measured) are agreed on by everyone, and ``reduction'' seems to be a
good name for this.  The term is not meant to include any additional
interpretational baggage.

As is well-known, time is not an observable in the technical sense in
quantum theory; it occurs as a parameter rather than an operator.
Thus, within quantum theory itself, we have no direct theory of the
measurement of time.  Yet perhaps this is not the right way of looking
at things.  Perhaps time is not merely a parameter, but another sort of
thing, in quantum theory.

\section{Happening in Quantum Theory}

Our knowledge and description of a quantum system thus comes from two
sources, the operators (whose dynamics are ultimately thought to be
governed by relativistic wave equations) and the state vector.
Indeed, from the strict point of view, nothing observable
ever {\em happens} except when a measurement is made.  All of the
intervening dynamics of the field variables simply serves to tell us
how the operators at one time are related to those at another.
Solving these dynamics allows us to predict the odds of getting
different measurements of different observables at a future time, but
we learn nothing definite until we make a measurement.
(This is one way of viewing the ``problem of time.'')

Two senses of ``time'' have appeared in this discussion.  The first
is that determined by the background space--time, which enables us to
say when one event precedes another, and by how much.  But the second
sense of time is that of ``objective happening,'' that is, the actual
things which mark an event as distinguished.  This 
``objective happening,'' is, in quantum
theory, expressed in the act of measurement.  We should take this
seriously, and aim to understand it.  It provides us with a whole new
perspective on the question of what time is.  

From this view, time, in the sense of objective happening, 
is {\em produced} by, indeed
{\em is} reduction of
the state vector.  This notion of time is as important as the relativistic one,
and the two notions of time should ultimately be united in a single theory.

\section{Quantum Field Theory and Measurement}\label{qftm}

I have so far discussed quantum theory in a general way, without
explicitly considering relativistic effects. 
To reconcile quantum theory with special relativity one passes to
quantum field theory.  There are several crucial new points which
occur.

First, an essential tenet of the theory is that all observable
operators be constructed from local quantum fields.  It is also
generally true that quantum fields are not operator-valued functions
on space--time, but operator-valued distributions.  This means that an
operator must generally be averaged over a space--time volume (against
a smooth weighting function); attempts to localize the averaging
process lead to more and more wildly fluctuating results, which are
manifestations of the divergent zero-point energy --- a point I will return to.

Suppose we have such an operator $Q$, which is the result of
integrating the quantum fields (perhaps sums of products of the
quantum fields) over a compact space--time volume $\tau$.  Using the
field equations, we may re-express this same operator in many
different ways, as integrals of the quantum fields against suitably
evolved weight functions supported on entirely different space--time
volumes (for instance, say $\tau '$).  Thus the operator $Q$, considered
simply as an operator, is not in any preferred way associated with a
single space--time volume.  In what sense, then, is the measurement of
$Q$ local to $\tau$?  When can it be known?

It is natural to assume that any actual measurement must take place
via some, as-yet incompletely understood, dynamics which is local in
space--time.  Thus we think that some sort of a device may be
constructed in a neighborhood of the volume $\tau$ effecting the
measurement of $Q$, or alternatively, a different sort of device might
be constructed on $\tau '$ effecting the same measurement.  (I am
using the term ``device'' for its intuitive appeal.  But it is not
meant to suggest that the devices need be manufactured or placed or
activated by conscious beings; it may well be that certain physical
configurations count as devices irrespective of how they are
attained.  I shall sketch a suggestion for this below.)

Let us now focus on one such measurement, by the device at $\tau$.
When can the result of the measurement be known?  
The dicta of quantum theory give us no answer to this.  On the other
hand, causality would seem to imply that the result of the measurement
can only be known to observers at events $p$ whose pasts include
$\tau$.  Thus the process of measurement, comprising reduction
together with learning the result, must take a finite time, determined
by causality.

While these considerations are very reasonable, there is something of
considerable importance for the shape of our conjectured
overall theory of time which has crept in:  
this is that the reduction only comes to be
objectively known at {\em points} $p$.   That is, even though the observable
measured might be considerably non-local (a field average over $\tau$,
for instance), and even though the reduction may necessarily take
place over a significant space--time volume, the actual objective
verification of it appears to require a considerable degree of
localization in space--time.

It is important to appreciate how strongly this statement depends on
some sort of a classical view, to good approximation, of space--time,
and how at odds it is with any quantum view of the fabric of
space--time itself.  If the fabric of space--time had any conventional
quantum character, one would think that, besides its usual
``position'' basis (knowing where events are)
there would be, for example a ``wave-number'' basis as
well (knowing the Fourier components describing extended
distributions); but it seems hard think of ascribing any definite
knowledge of the reduction of a state to a particular wave-number.
This is an indication that space--time enters the reduction
process in an essentially
different way from quantum fields --- and this will have implications
for reconciling quantum theory with general relativity, an issue which
I will return to below.

\section{Renormalization, Stress--Energy and Time}

A natural approach to understanding time in quantum theory is via
its classically conjugate variable, energy.  While a great deal of
interesting work has been done, for instance, on the time--energy
uncertainty relation via quantum-mechanical thought experiments, I
want to explore something essentially different here, relativistic
effects for which one needs quantum field theory and renormalization.
What I shall argue here is that there is good reason to think that
time plays a more profound role than has been considered in
renormalization and in the meshing of quantum physics to classical
space--time.
I shall consider here linear quantum field theories.

\subsection{Definition of the Stress--Energy Operator}

The starting-point for relativistic
treatments of the matter-content of a system is the stress--energy,
and in quantum field theory this becomes the stress--energy operator.
In linear quantum theories, it has formally the same appearance as in
the corresponding classical theories, but it is divergent (even as an
operator-valued distribution) and must be renormalized.  That
renormalization is accomplished by {\em normal-ordering}, that is,
re-writing the factors so that no annihilation operator precedes its
corresponding creation operator.  At a formal level, the
normal-ordering subtracts a c-number divergent term, the vacuum
stress--energy, from the formal one defined by the local fields:
\begin{equation}
 T_{ab}^{\rm renormalized}=T_{ab}^{\rm local\ fields}-T_{ab}^{\rm vacuum}\,
.
\end{equation}
The normal-ordering on which this is based is very much a temporal
concept, for the creation and annihilation operators are defined as
the negative- and positive-frequency parts of the fields.  

What happens in curved space--time?  The construction of the quantum
field theory is quite similar; indeed, if one can determine which
modes of the field count as positive- and which as negative-frequency,
one can mimic the entire special-relativistic construction.  This is
usually expressed in terms of choosing a complex structure $J$ 
(with $J^2=-1$) on a
suitable space $\Gamma$ of classical solutions of the field equations,
the positive- and negative-frequency modes corresponding to the two
eigenspaces of $J$.

The choice of $J$ is not unique, however, and so one really has a
family of possible quantizations, indexed by the acceptable $J$'s.
One requires that the $J$'s lead to $n$-point functions whose
dominant ultra-violet asymptotics are the same as for fields in
Minkowski space; this considerably restricts $J$, to {\em
Hadamard} representations.  However, some ambiguity remains.

This ambiguity can be fixed in certain circumstances, such as in a
stationary space--time with suitable
asymptotics.  However, the $J$'s associated with two different
stationary regimes will in general be different, being related by a
{\em Bogoliubov transformation}.  This means that there is no absolute
sense to a creation or an annihilation operator, and no absolute sense
to the quantum vacuum.  Indeed, it is just this last fact which gives
rise to the possibility of an initially vacuum state becoming, without
any change, an occupied state --- the sense of what ``vacuum'' and
``occupied'' are has altered.

The definition of $T^{\rm renormalized}_{ab}$  thus depends on the definition of
normal-ordering, which in turn depends on $J$, which is ambiguous because of the
lack of a preferred time coordinate. Thus in general it is possible for the
definition of $T^{\rm renormalized}_{ab}$  to be ambiguous, although one might
hope that in (for instance) stationary regimes a preferred definition of $T^{\rm
renormalized}_{ab}$ would exist.   It is known that, if one requires $T^{\rm
renormalized}_{ab}$  to be conserved ($\nabla ^a T_{ab}=0$), then choices of
$T^{\rm renormalized}_{ab}$ do  exist which are unique up to a possible
conserved c-number addition.

I think it is generally considered that this construction of the
stress--energy is essentially correct, and that probably if we are
clever enough we will be able to fix the c-number ambiguity.  However,
there are several points at which I think this view may be mistaken.

\subsection{Negative Energies and Measurement}

It should first be appreciated that the stress--energy operator, even as an
operator-valued distribution, is a rather singular object no matter how the
renormalization is done.
One would think, by analogy with classical theories, that on a Cauchy
surface $\Sigma$ the Hamiltonian for evolution along the
future-directed timelike vector field $\xi ^a$ would be
\begin{equation}
 H(\Sigma , \xi )=\int T_{ab}^{\rm renormalized}\, \xi ^a\,
\rmd\Sigma ^b\, .
\end{equation}
This turns out to be true in a weak sense, but $H(\Sigma ,\xi ^a)$ is
in generic circumstances rather pathological~(Helfer 1996).
It is (unless $\xi ^a$ is a Killing vector) 
not self-adjoint (so it does not have a well-defined spectral resolution and
cannot directly be an observable) and it is generically unbounded
below (no matter {\em which} allowable c-number contribution to the
renormalization is chosen).  
If the integral above is replaced by an average over a space--time
volume close to $\Sigma$, then one can avoid these problems,
but as the volume over which the average is taken approaches $\Sigma$,
the lower bounds tend to $-\infty$.  (This occurs
even in special relativity, if $\xi ^a$ is perturbed by any finite
amount from a Killing field.)
Thus
$T_{ab}^{\rm renormalized}$ cannot be a conventional measure of
energy-content, for it would predict an unstable theory.  

We are led to conclude that energy cannot be well-localized in time 
in quantum
field theory.
While this behavior is quite different from the non-relativistic
time--energy uncertainty relations, it is bound up with them.  I have
argued elsewhere (Helfer~1998) that it is in effect {\em
symptomatic} of the problem of giving an objective meaning to
space--time geometry in the absence of quantum measurement
considerations,
and that any attempt to
verify negative energies' existence directly would likely require a measuring
device which would give a net positive local energy --- underscoring
the lack of objective reality which can be ascribed to such
situations, in the absence of quantum measurements.

\subsection{Time and Nonlocality of Renormalization}

The ambiguity surrounding the definition of the stress--energy has obscured an
important issue.  It is highly likely that the renormalization prescription
forces the stress--energy to be determined by considerations which are {\em
nonlocal in time}. 
Indeed, in simple cases where one can renormalize in a preferred way, one sees
this directly.  The complex structure $J$ distinguishes positive from negative
frequencies, and this requires a temporal averaging to define.  (And the various
renormalization procedures used in relativistic quantum field theory are also
nonlocal, although the divergent terms are often local.)
I should make it clear that this possibility is not usually considered, and
indeed, as a way of trying to fix the ambiguity many physicists actually make
the opposite assumption (that is, that the renormalization should be local).
However, the physical underpinnings of renormalization have very much to do with
averaging and considerations of different scales, and so my view is that the
more likely possibility is that these nonlocal averages are real.

For linear quantum field theories, these questions about the correct
renormalization prescription only contribute finite c-number modifications to
the stress--energy.  Thus one might think that even if the nonlocal averaging
prescription is correct, it only contributes minor technical modifications. 
However, the nonlocality involved changes the theory in
a fundamental way.

The nonlocality would mean that the stress--energy, and hence the
Hamiltonian operators generating evolution, cannot be locally known.  Thus the
energy--momentum-content of a region can at best be inferred only after the fact
in light of subsequent developments.  I would suggest that this is a new sort of
time--energy uncertainty relation.  It has in fact a dual character, according
to the two concepts of time I have considered:  from the usual relativistic
point of view, it restricts how well the energy can be known locally; but from
the point of view of time as reduction, it means that a reduction measuring such
a Hamiltonian must occur over a larger volume of space--time than that
indicated
by the local operators which go into the stress--energy.  (That is, longer than
that deduced from general principles in section~\ref{qftm}.)

There may be the possibility of observing the consequences of these effects
interferometrically. 
The c-number
character of the vacuum term is not absolute; we may imagine quantum operators
coupling to it:  for instance, if the field interacts with boundaries via
Casimir effects, if the precise location of the boundaries is switched according
to the quantum state of other operators.

\subsection{Quantum General Relativity?}

I have already indicated that while it is crucial to integrate the
relativistic notion of time with the quantum concept of reduction,
there are indications that doing so will treat space--time and quantum
theory in different ways.  I pointed out that objective verification
that reduction has occurred appears to be an inherently {\em local}
concept, which is simply difficult to reconcile with the usual view of
quantum theory as being largely independent of notions of localization
in space--time.  

I also pointed out that measures of energy given by the stress--energy
operator are unbounded below, and this suggests that it will be very
hard to quantize Einstein's equation
\begin{equation} 
G_{ab}=-8\pi G T_{ab}\
\end{equation}
by somehow promoting both sides to renormalized quantum operators and
have a physically sensible theory.  The unboundedness-below of
$T_{ab}^{\rm renormalized}$ suggests a gross unphysical instability.

And I pointed out that the renormalization of $T_{ab}$ involved the
subtraction of a divergent vacuum term $T_{ab}^{\rm vacuum}$.  In the
context of linear quantum field theory in curved space--time, this
term was a c-number, but if the metric is somehow promoted to a
quantum operator, the vacuum term, which depends very much on the
metric, becomes a fluctuating q-number which, as we have seen,
is moreover likely to be nonlocal in time.  This nonlocality moves us
beyond the realm of conventional quantum theory.

Thus it seems to me that, while it is absolutely necessary to modify
general relativity in some sense to meld it with quantum theory and
reduction, there are a number of strong points arguing that the
correct modification will not be to simply quantize the metric as
another field.  Something will need to be done which to some degree
preserves the locality of relativity, and it is unlikely that the
quantum stress--energy tensors can be in any direct sense considered as
sources for whatever the correct extension of general relativity to
quantum theory is.

\section{What Determines Reduction?}

The question of what determines the reduction of the state vector is, as I
indicated earlier, a key element which is simply missing from our physical
understanding at present.  Work on this question is still in very preliminary
stages.  Here I shall indicate not a particular solution, but what appear to me
to be some fruitful avenues for its exploration, together with some
consequences.

A prescription for reduction must determine both when it is likely to occur, and
what operator is likely to be measured.  The fact that the world does not
spontaneously reduce to eigenstates of bizarre operators is a considerable
restriction on the choice.  Also the fact that there is so much freedom in
placing the ``classical/quantum cut'' suggests that by and large the operators
which are measured 
very nearly commute (on the subspaces of Hilbert space in which the
states are known to lie) with the Hamiltonians.

These considerations, together with a desire for economy, suggest that the
mechanism of reduction should be somehow coded in the Hamiltonians themselves,
or in related structure, and should also depend on the subspace of Hilbert space
in which the state vector is already known to lie.

While I believe that reduction is likely to occur via non-gravitational forces
as well as gravitational ones
(presumably, in most physics laboratories most reduction is due somehow to
electromagnetic interactions), taking up this issue would lead to consideration
of nonlinear field theories.  (Preliminary work does indicate that there
is indeed scope for applying the ideas here to such theories.)
So I will confine my
remarks to linear quantum fields in curved space--time.  (Similar
ideas hold for linear fields in the
presence of external potentials.)

We have seen that the Hamiltonians of linear quantum fields
\begin{equation}
 H(\Sigma ,\xi )=\int _\Sigma T_{ab}^{\rm renormalized}\, \xi ^a\rmd \Sigma
^b
\end{equation}
are really rather
singular objects, being generically unbounded below and not self-adjoint, 
and that what we really should consider are
temporally-averaged Hamiltonians, which would have the general form
\begin{equation}
 H(\tau ,\xi ^{ab}) =\int _\tau T_{ab}^{\rm renormalized}\, \xi
^{ab}\rmd\tau\, ,
\end{equation}
where $\tau$ is a space--time volume in the neighborhood of a Cauchy surface
$\Sigma$, and now $\xi ^{ab}$ encodes both the vector field determining
evolution and the normals to the hypersurfaces being averaged over.  These
averaged Hamiltonians are self-adjoint, and for suitable $\xi ^{ab}$ they are
bounded below, although as the averaging narrows to a particular $\Sigma$ the
individual lower bounds diverge to $-\infty$.  

It seems plausible that what should determine reduction would be a tendency to
seek situations in which these Hamiltonians were relatively non-singular.  Thus
it is possible that (say) reduction becomes likely for those space--time volumes
$\tau$ and those $\xi ^{ab}$ for which the lower bound approaches zero.  In this
view, reduction over very short times (and the attendant
high quantum fluctuations) would be highly suppressed, and also reduction along
very ``twisted'' $\xi ^{ab}$ would be highly suppressed; in fact, the reduction
would tend to select, as nearly as possible, tensors $\xi ^{ab}=\xi ^a t^b$,
where $t^a$ would be a normal to the surfaces $\Sigma$ and $\xi ^a$ would
approximate a timelike Killing vector.  It would also be expected that the
neighborhood of a Cauchy surface could well develop different domains in
which reduction proceeded independently, according as in each domain the
criterion for reduction became increasingly likely to be satisfied.

If something like this can be achieved, it would {\em of itself} provide a
definition of a ``cosmic time vector'' to good approximation, that is, the
temporal Killing vector $\xi ^a$ which is approximately a component of $\xi
^{ab}$.

\section{Conclusion}

I have suggested that reduction of the state vector should be
considered a sort of time, as important as the conventional
relativistic one, and that it is essential to reconcile these notions.
From the point of quantum theory, a key question --- whether one
accepts the rest of the arguments in this paper or not --- is, Under
what circumstances does a measurement take place?  

I have argued that it is likely that the information signalling
reduction is somehow largely bound up in the stress--energy operator,
and that a correct resolution of that operator's apparently
pathological properties might be that they point the way to reduction.
A rough indication was given of a general mechanism which might
determine both reduction and (to reasonable approximation) a cosmic
flow of time.

The problem of reconciling the quantum and the relativistic notions of
time gives a point of view on the problem of reconciling quantum
theory and general relativity --- what is usually called quantum
gravity.  While this issue is very important and will, in my view,
involve some modification of general relativity, it is not at all
clear that that modification will be a quantization of the metric in a
conventional sense; there are a number of indications to the contrary.

The ideas here are of course only tentative templates for more
detailed explorations.

\section*{References}

\noindent Helfer, A. D. (1996) Class. Quantum Grav. {\bf 13}, L129.

\noindent Helfer, A. D. (1998) Class. Quantum. Grav. {\bf 15}, 1169.

\noindent Leggett, A. J. (2002) J. Phys.:  Condens. Matter {\bf 14} R415.

\noindent Peacock, J. A. (1999) Cosmological Physics (Cambridge: University
Press).

\end{document}